\documentclass[a4paper,11pt]{article}
\usepackage{jcappub} 
\usepackage{lineno}
\usepackage{placeins}

\title{\boldmath Environmental Dependence of Satellite Planarity and Orbital-Pole Coherence around Milky Way Analogues in a Hydrodynamical Cosmological Simulation}

\author[a]{Peng Wang,*}
\author[b]{Yu Luo,*}
\author[c]{Jiang Chang,*}
\author[d]{Xi Kang,*}
\author[e,f]{Shi Shao,*}

\affiliation[a]{Shanghai Astronomical Observatory, Chinese Academy of Sciences, Nandan Road 80, Shanghai 200030, People's Republic of China.}
\affiliation[b]{Department of Physics, School of Physics and Electronics, Hunan Normal University, Changsha 410081, People’s Republic of China.}
\affiliation[c]{Purple Mountain Observatory, Chinese Academy of Sciences, No.10 Yuan Hua Road, 210034 Nanjing, People’s Republic of China}
\affiliation[d]{Institute of Astronomy, School of physics, Zhejiang University, Hangzhou 310027, People's Republic of China.}
\affiliation[e]{Institute of Astronomy and Physics, Inner Mongolia University, Hohhot, 010021, China.}
\affiliation[f]{National Astronomical Observatories, Chinese Academy of Sciences, Beijing, 100101, China.}

\emailAdd{pwang@shao.ac.cn, luoyupmo@gmail.com, changjiang@pmo.ac.cn, kangxi@zju.edu.cn, shaoshi@nao.cas.cn}

\abstract{We examine whether the unusual spatial flattening and orbital-plane coherence of the Milky Way's classical satellites are connected to the Galaxy's location in the cosmic web. We compare the 11 classical satellites with the 11 most massive satellites of 11,922 Milky Way-mass hosts in the TNG300-1 hydrodynamical simulation. We measure flattening with both the simple inertia tensor (SIT), which retains radial information, and the reduced inertia tensor (RIT), which isolates angular anisotropy. Orbital-plane coherence is characterized by the minimum opening angle containing the axes of the eight most coplanar satellite orbits. For the Milky Way, we find $(c/a,b/a)=(0.183,0.558)$ for SIT and $(0.364,0.736)$ for RIT, together with an eight-orbit opening angle of $21.9^\circ$. The corresponding fractions of simulated systems at least as extreme are $4.94\%$, $12.28\%$, and $5.04\%$, respectively. We describe the environment using the velocity-shear tensor smoothed on $2\,h^{-1}\mathrm{Mpc}$ and compare the full sample with Local Group-like pairs, systems whose shear eigenvalues match those reconstructed for the Local Group, and hosts in knots, filaments, sheets, and voids. Among the well-sampled environmental subsets, the shape and opening-angle distributions differ only weakly. Although the cosmic web imprints preferred orientations on host spins and satellite orbital-plane axes, the Milky Way combines orientations that are uncommon in the simulation. We conclude that the present-day cosmic-web environment alone does not substantially increase the occurrence of Milky Way-like satellite configurations. This result does not rule out an origin in anisotropic accretion, since neither the present-day web class nor its shear eigenvalues fully encodes the assembly history of an individual satellite system.}

\begin{document}
\maketitle
\flushbottom

\section{Introduction}
\label{sec:intro}

Dwarf satellite galaxies are sensitive probes of galaxy formation and the small-scale distribution of dark matter. Deep wide-field surveys, including the Sloan Digital Sky Survey, the Dark Energy Survey, and Pan-STARRS, have greatly enlarged the census of low-luminosity satellites in the nearby Universe \citep{2000AJ....120.1579Y,2016MNRAS.460.1270D,2016arXiv161205560C}. Their luminosities, internal dynamics, and spatial distributions therefore provide stringent tests of the $\Lambda$ cold dark matter ($\Lambda$CDM) model \citep{1999ApJ...522...82K,1999ApJ...524L..19M,2011MNRAS.415L..40B,2017ARA&A..55..343B}.

A prominent example is the flattened distribution of satellites around massive host galaxies. The classical Milky Way (MW) satellites delineate the Vast Polar Structure \citep{2012MNRAS.423.1109P}; a subset of the M31 satellites lies in a thin structure with correlated line-of-sight velocities \citep{2013Natur.493...62I}; and a comparable configuration has been reported around Centaurus~A \citep{2018Sci...359..534M,2021A&A...645L...5M}. It is important, however, to distinguish spatial flattening, orbital-plane coherence, and coherent rotation. A thin configuration observed at one epoch need not be dynamically long-lived, while a cluster of orbital-plane axes may contain both co- and counter-rotating members.

The frequency and interpretation of satellite planes within $\Lambda$CDM remain under debate. Some comparisons with simulations find MW-like spatial and kinematic configurations to be rare \citep{2021NatAs...5.1185P,2021Galax...9...66P}. Others stress their sensitivity to sample selection, the chosen statistic, the radial distribution of the satellites, and a posteriori choices \citep{2015MNRAS.452.3838C,2017ApJ...843...62M,2021NatAs...5.1188B,2023NatAs...7..481S}. Numerical studies further indicate that a spatial plane may be short-lived even when some of its orbital planes remain correlated for longer \citep{2019MNRAS.488.1166S,2026Sawala}. Meaningful comparisons must therefore use identical definitions for observed and simulated systems and keep spatial anisotropy distinct from orbital coherence.

The cosmic web provides a natural setting in which such correlations might arise. Matter and subhaloes are accreted anisotropically through sheets and filaments, and correlated or group infall may leave preferred directions in the satellite population\citep{2011MNRAS.411.1525L,2018ApJ...859..115W,2020ApJ...900..129W,2021ApJ...914...78W}. Reconstructions of the local velocity field have suggested that several nearby satellite planes are oriented relative to the principal axes of the surrounding velocity shear \citep{2015MNRAS.452.1052L}. This connection can be studied with constrained simulations of the Local Group, such as HESTIA \citep{2020MNRAS.498.2968L,2022MNRAS.516.4576D}, or statistically with a large cosmological volume containing many MW-mass hosts in a range of environments.

Here we adopt the latter approach and use TNG300-1 from the IllustrisTNG suite \citep{2018MNRAS.480.5113M,2018MNRAS.477.1206N,2018MNRAS.475..624N,2018MNRAS.475..648P,2018MNRAS.475..676S}. We measure the spatial shape and orbital-plane coherence of the 11 classical MW satellites and of equally sized satellite samples around simulated MW analogues. We then test whether a Local Group-like companion or an MW-like velocity-shear environment alters the frequency of extreme satellite configurations. We also compare the satellite orbital-plane axis and host spin with the principal axes of the cosmic web. The tail fractions quoted below are descriptive; because the statistics and selection were not specified independently of the observed MW system, we do not interpret them as model-exclusion probabilities.

\section{Data and Methodology}
\label{sec:data_method}

We compare the phase-space distribution of the 11 classical MW satellites with samples of the same size around simulated MW-mass hosts. We make this choice because the classical satellites have comparatively well-constrained kinematics, while the 11 most massive satellites around each simulated host can be robustly resolved at the adopted numerical resolution. Holding the sample size fixed permits a direct comparison and avoids introducing a dependence on satellite number into the shape and orbital-pole statistics. We first describe the observational and simulated samples, followed by the measures of spatial shape, orbital-pole coherence, and cosmic-web environment.

\subsection{Observational Data}

We adopt the phase-space compilation of ref.~\cite{2019MNRAS.488.1166S}. Sky positions, distances, and line-of-sight velocities are taken from ref.~\cite{2012AJ....144....4M}. The proper motions are based on Gaia DR2 \citep{2018A&A...616A...1G}, supplemented by Hubble Space Telescope measurements for Leo~I and Leo~II \citep{2013ApJ...768..139S,2016AJ....152..166P}. We transform the observables to a right-handed Galactocentric Cartesian frame following ref.~\cite{2015MNRAS.452.3838C}. The resulting phase-space coordinates are listed in Table~\ref{table:table1}. Throughout this work, we use the tabulated central values without propagating measurement uncertainties. The observational measurements should thus be regarded as a fixed benchmark rather than a posterior distribution.

\begin{table*}
\caption{Phase-space coordinates and orbital-pole directions of the 11 classical MW satellites. Columns 2--4 give Galactocentric Cartesian positions $(x,y,z)$ in kpc, columns 5--7 give Galactocentric velocities $(v_x,v_y,v_z)$ in km\,s$^{-1}$, and the final two columns list the Galactic longitude and latitude of the orbital pole, $(l_{\rm pole},b_{\rm pole})$, in degrees.}
\begin{center}  
\begin{tabular}{c|ccccccccc}
\hline
Name           & $x$ & $y$ & $z$ & $\upsilon_{x}$ & $\upsilon_{y}$ & $\upsilon_{z}$ & $l_{\rm pole}$ & $b_{\rm pole}$ \\
               & (kpc) & (kpc) & (kpc) & (km/s) & (km/s) & (km/s) & ($^{\circ}$) & ($^{\circ}$) \\
\hline
Sagittarius    &  11.566 &   1.684 &  -4.368 &  233.142 &  -17.079 &  206.379 & 274.6 &  -9.8 \\
LMC            &  -0.371 & -28.281 & -18.607 &  -41.666 & -217.598 &  222.248 & 175.3 &  -6.0 \\
SMC            &  11.186 & -26.066 & -30.261 &    7.131 & -184.215 &  171.660 & 192.0 & -10.3 \\
Draco          &  -2.960 &  42.162 &  29.252 &   47.278 &   3.548  & -150.196 & 171.7 & -17.2 \\
Ursa Minor     & -15.049 &  35.218 &  36.207 &   -1.393 &  46.395  & -147.149 & 198.3 &  -5.1 \\
Sculptor       &  -3.542 &  -6.567 & -57.745 &   17.120 & 176.172  & -96.186  & 353.0 &  -2.7 \\
Sextans        & -24.827 & -38.538 &  39.184 & -241.126 &  90.287  &  58.395  & 234.1 & -49.4 \\
Carina         & -16.928 & -64.835 & -26.859 &  -40.741 & -47.574  & 160.479  & 161.9 &  -8.5 \\
Fornax         & -27.857 & -34.385 & -90.702 &   45.455 & -132.824 &  83.749  & 186.8 &  19.3 \\
Leo II         & -52.301 & -39.455 & 145.416 &  -39.401 & 118.873  &  39.733  & 191.0 & -22.0 \\
Leo I          & -83.669 & -80.830 & 129.719 & -166.689 & -35.307  &  95.468  & 257.0 & -36.9 \\
\hline
\end{tabular}
\end{center}
\label{table:table1}
\end{table*}

\subsection{Simulation Data}

We analyse the $z=0$ output of TNG300-1, the largest full-physics magnetohydrodynamical simulation in the IllustrisTNG suite \citep{2018MNRAS.475..676S,2018MNRAS.475..648P,2018MNRAS.477.1206N,2018MNRAS.475..624N,2018MNRAS.480.5113M}. The simulation was run with the moving-mesh code \textsc{arepo} \citep{2010MNRAS.401..791S} in a periodic volume of side length $205\,h^{-1}\mathrm{Mpc}$ ($302.6\,\mathrm{Mpc}$). It contains $2500^3$ dark-matter particles and initially the same number of gas cells, with dark-matter and target baryonic mass resolutions of $5.9\times10^7\,M_\odot$ and $1.1\times10^7\,M_\odot$, respectively. The adopted cosmological parameters, $\Omega_m=0.3089$, $\Omega_\Lambda=0.6911$, $\Omega_b=0.0486$, $\sigma_8=0.8159$, $n_s=0.9667$, and $h=0.6774$, are consistent with the Planck 2015 results \citep{2016A&A...594A..13P}.

Dark-matter haloes are identified using a friends-of-friends algorithm with a linking length equal to 0.2 times the mean interparticle separation \citep{1985ApJ...292..371D}. The gravitationally bound substructures are then located with \textsc{subfind} \citep{2001MNRAS.328..726S}, including the baryonic components as described in ref.~\cite{2009MNRAS.399..497D}. We define the most massive substructure in each friends-of-friends group as the central subhalo and measure all positions relative to its potential minimum.

\subsection{Milky Way Analogues}

MW analogues are selected according to
\begin{itemize}
    \item $0.5 \leq M_{200} / (10^{12}\,M_\odot) \leq 2$,
    \item $N_{\rm sate} \geq 11$,
\end{itemize}
Here $M_{200}$ is the mass enclosed within a radius whose mean density is 200 times the critical density, and $N_{\rm sate}$ is the number of resolved, bound satellite subhaloes. For each selected host, we rank the satellites by dark matter mass and retain the 11 most massive, matching the classical MW sample. These criteria yield a fiducial catalogue of 11,922 hosts.

\subsection{Satellite Spatial Plane}

We characterize the three-dimensional satellite distribution with two shape tensors,
\begin{align}
I^{\rm SIT}_{ij} &= \sum_{k=1}^{N} x_{i,k}x_{j,k}, \\
I^{\rm RIT}_{ij} &= \sum_{k=1}^{N}\frac{x_{i,k}x_{j,k}}{R_k^2},
\label{equ:plane}
\end{align}
where $\boldsymbol{x}_k$ is the position of satellite $k$ relative to the central galaxy, $R_k=|\boldsymbol{x}_k|$, and $i,j\in\{x,y,z\}$. In the simple inertia tensor (SIT), the contribution of a satellite scales as $R_k^2$, making the inferred shape sensitive to the radial distribution and, in particular, to distant satellites. The reduced inertia tensor (RIT) instead projects all positions onto the unit sphere. Each satellite then carries the same total weight, and the tensor retains only the angular anisotropy \citep{2005ApJ...627..647B,2023NatAs...7..481S}.

For both tensors, we order the eigenvalues as $\lambda_{s1}\geq\lambda_{s2}\geq\lambda_{s3}$ and denote the corresponding unit eigenvectors by $\boldsymbol{e}_a$, $\boldsymbol{e}_b$, and $\boldsymbol{e}_c$. Up to a common normalization, the principal-axis lengths are $a=\sqrt{\lambda_{s1}}$, $b=\sqrt{\lambda_{s2}}$, and $c=\sqrt{\lambda_{s3}}$. We characterize the shape by $b/a$ and $c/a$, with smaller $c/a$ indicating a flatter configuration. SIT and RIT address different aspects of the distribution---the full spatial anisotropy and the angular anisotropy, respectively---and we therefore report both without assigning either one a privileged status.

\subsection{Satellite Orbital Clustering}
\label{sec:opening_angle}

Following ref.~\cite{2019MNRAS.488.1166S}, we quantify orbital coplanarity using the minimum opening angle $\alpha_N$. For each satellite $j$, we calculate its unit orbital-angular-momentum vector $\hat{\boldsymbol{L}}_j$. We then draw $10^5$ trial axes $\{\hat{\boldsymbol{n}}_i\}$ uniformly over the sphere and compute
\[
\alpha_{ij} = \arccos\!\left(\big|\hat{\boldsymbol{n}}_i \cdot \hat{\boldsymbol{L}}_j\big|\right),
\]
where the absolute value treats $\hat{\boldsymbol{L}}_j$ and $-\hat{\boldsymbol{L}}_j$ as equivalent. The statistic thus measures a common orbital plane while allowing counter-rotating members; by itself, it does not measure a common sense of rotation. For each trial axis, we sort $\{\alpha_{ij}\}$ and retain the largest among the $N$ smallest angles,
\[
\alpha_{N,i} \equiv \max\{\alpha_{ij}\ \text{of the $N$ smallest in } j\}.
\]
The minimum opening angle is
\[
\alpha_N \equiv \min_i \alpha_{N,i}.
\]
A smaller $\alpha_N$ indicates tighter clustering of $N$ orbital-plane axes. For the MW, the marked increase from $N=8$ to $N=9$ motivates our choice of $N=8$; in the figures, we denote $\alpha_8$ by $\phi_8$. We reflect each selected pole into the hemisphere centred on the best-fitting axis, average the eight resulting unit vectors, and denote the normalized mean by $\hat{\boldsymbol{J}}^{8}_{\rm mean}$. This sign convention is required when interpreting alignments involving an otherwise unoriented orbital-plane axis.

\subsection{Cosmic Web}

We characterize the cosmic web using the velocity-shear tensor \citep{2012MNRAS.425.2049H}. The density and mass-weighted peculiar-velocity fields are assigned to a $256^3$ Cartesian mesh by cloud-in-cell interpolation. We smooth the velocity field with a Gaussian filter of fiducial radius $R_s=2\,h^{-1}\mathrm{Mpc}$ and evaluate the derivatives in Fourier space. The resulting dimensionless symmetric tensor is
\begin{equation}
\Sigma_{ij} = -\frac{1}{2H_0} \left( \frac{\partial v_i}{\partial r_j} + \frac{\partial v_j}{\partial r_i} \right),
\label{equ:shear}
\end{equation}
where $\boldsymbol{v}$ is the peculiar-velocity field at $z=0$. Positive eigenvalues indicate local compression, whereas negative values indicate expansion.

Following ref.~\cite{2020NewA...8001405W}, we interpolate the six independent tensor components to each halo position and then diagonalize the interpolated tensor; we do not interpolate precomputed eigenvalues or eigenvectors. We order the eigenvalues as $\lambda_1\geq\lambda_2\geq\lambda_3$, with corresponding orthonormal eigenvectors $\boldsymbol{e}_1$, $\boldsymbol{e}_2$, and $\boldsymbol{e}_3$. With this convention, $\boldsymbol{e}_1$ is the fastest-collapse direction, while $\boldsymbol{e}_3$ is the slowest-collapse (or fastest-expansion) direction.

We classify the web using a zero eigenvalue threshold. Regions with zero, one, two, or three positive eigenvalues are labelled as voids, sheets (or walls), filaments, and knots, respectively. We apply the same threshold and smoothing scale to all hosts. In addition, we define an ``LSS-matched'' sample by requiring each of the three eigenvalues to fall within the corresponding $3\sigma$ interval reconstructed for the Local Group in ref.~\cite{2015MNRAS.452.1052L}. This selection leaves 2,049 of the 11,922 fiducial hosts.

\subsection{Alignment Angles}

We compare the satellite minor axis $\boldsymbol{e}_c$, the folded mean orbital-pole direction $\hat{\boldsymbol{J}}^8_{\rm mean}$, the unit host-spin vector $\hat{\boldsymbol{J}}_{\rm host}$, and the three velocity-shear eigenvectors. We define the following alignment cosines:
\begin{align}
\mu_{cJ} &= |\boldsymbol{e}_c\cdot\hat{\boldsymbol{J}}_{\rm host}|, \\
\mu_{8J} &= |\hat{\boldsymbol{J}}^8_{\rm mean}\cdot\hat{\boldsymbol{J}}_{\rm host}|, \\
\mu_{Ji} &= |\hat{\boldsymbol{J}}_{\rm host}\cdot\boldsymbol{e}_i|, \\
\mu_{8i} &= |\hat{\boldsymbol{J}}^8_{\rm mean}\cdot\boldsymbol{e}_i|.
\end{align}
All four quantities lie in $[0,1]$, with unity denoting parallel axes and zero denoting perpendicular axes. The absolute values remove the arbitrary signs of the tensor eigenvectors and orbital-plane normals. Consequently, $\mu_{8J}$ measures the alignment between the satellite orbital plane and the host equatorial plane, rather than a signed co-rotation fraction.

\section{Results}
\label{sec:results}

We begin by measuring the spatial flattening and orbital-plane coherence of the MW satellite system, which sets the observational benchmark for the simulation. We then examine whether Local Group-scale pairing or the larger-scale velocity-shear environment affects the frequency and orientation of comparable configurations.

\subsection{Satellite Planarity and Orbital-Pole Coherence of the Milky Way}

\begin{table*}[htbp]
\caption{Spatial-shape and orbital-plane statistics for the 11 classical MW satellites. Each absolute cosine is followed in parentheses by the corresponding angle. The orbital-plane axis $\hat{\boldsymbol{J}}^8_{\rm mean}$ is calculated after reflecting the counter-rotating pole into the common hemisphere.}
\begin{center}
\begin{tabular}{lccc}
\hline\hline
Spatial measure & $c/a$ & $b/a$ & $\mu_{cJ}$ \\ \hline
SIT & 0.183 & 0.558 & $0.211\;(77.8^\circ)$ \\
RIT & 0.364 & 0.736 & $0.360\;(68.9^\circ)$ \\
\hline
Orbital measure & $\phi_8$ & $\hat{\boldsymbol{J}}^8_{\rm mean}$ & $\mu_{8J}$ \\ \hline
 & $21.9^\circ$ & $(-0.994,-0.021,-0.106)$ & $0.11\;(83.9^\circ)$ \\
\hline\hline
\end{tabular}
\end{center}
\label{table:table2}
\end{table*}


\begin{figure}[htbp]
\centering
\includegraphics[width=0.98\textwidth]{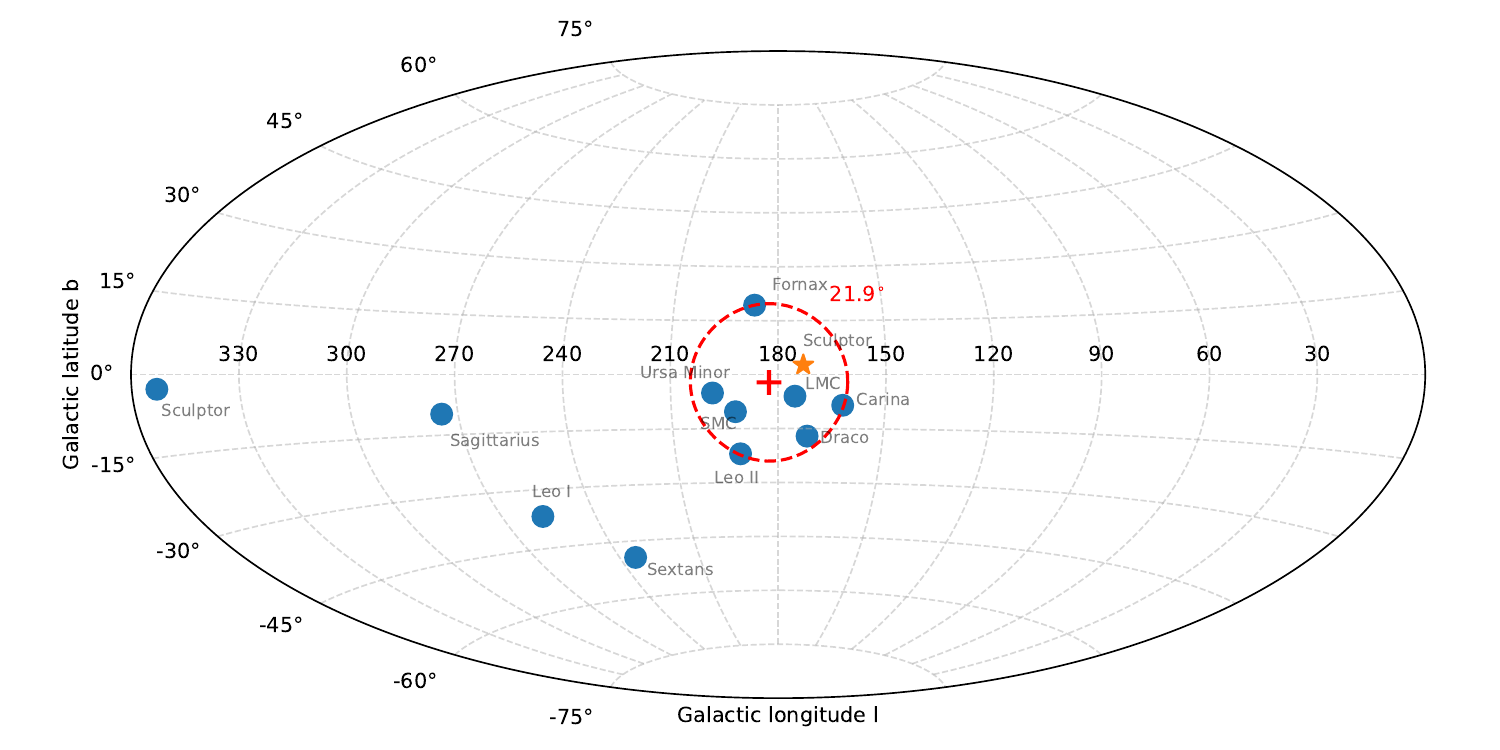}
\caption{Aitoff equal-area projection of the orbital poles of the 11 classical MW satellites. The blue points show the measured poles in Galactic coordinates. The red cross marks the best-fitting orbital-plane axis, and the dashed circle encloses the $21.9^\circ$ opening angle of the eight most coplanar orbits. Sculptor lies near the antipode of the red cross: it shares the orbital plane but rotates in the opposite sense.}
\label{fig:f2}
\end{figure}

Using the definitions introduced in Section~\ref{sec:data_method}, we measure the spatial-shape and orbital-plane properties of the 11 classical MW satellites, as summarized in Table~\ref{table:table2}. For SIT, the MW satellites have $(c/a,b/a)=(0.183,0.558)$. For RIT, the ratios increase to $(0.364,0.736)$ (right panel), indicating that the angular distribution is less extreme than the full three-dimensional distribution. This difference is readily understood: the distant satellites Leo~I and Leo~II have considerable leverage in SIT, whereas every satellite has the same total weight in RIT. The dependence on tensor definition therefore carries information about the MW's uneven radial sampling rather than representing a fitting uncertainty \citep{2023NatAs...7..481S}. Despite the difference in axis ratios, the minor-axis directions obtained with SIT and RIT differ by only $9.8^\circ$. Their alignment cosines with the Galactic spin are $\mu_{cJ}=0.21$ and $0.36$, corresponding to misalignment angles of $78^\circ$ and $69^\circ$, respectively. Both definitions therefore identify a broadly polar satellite configuration relative to the MW disc, although the RIT minor axis is somewhat less perpendicular to the Galactic spin \citep{2013MNRAS.435.2116P,2016MNRAS.460.3772S}.

The eight most coplanar satellite orbits have $\phi_8=21.9^\circ$ and a folded mean direction $\hat{\boldsymbol{J}}^8_{\rm mean}=(-0.994,-0.021,-0.106)$. Figure~\ref{fig:f2} also illustrates an important distinction: seven measured poles cluster on one side of the sphere, whereas Sculptor lies near the antipode. Thus, eight satellites share approximately the same orbital plane, but only seven have the displayed sense of rotation. The orbital-plane axis is almost perpendicular to the MW spin, with $\mu_{8J}=0.11$ ($84^\circ$). The spatial plane and the orbital-plane axis both describe a broadly polar configuration, although they are not equivalent measurements.

\subsection{Satellite Planarity and Orbital-Pole Coherence in Milky Way Analogues}

We next apply the same statistics to the 11 most massive satellites of each simulated host. We classify a system as at least as flattened as the MW only if both $b/a$ and $c/a$ are no larger than the corresponding MW values measured with the same tensor.

\begin{figure}[htbp]
\centering
\includegraphics[width=0.98\textwidth]{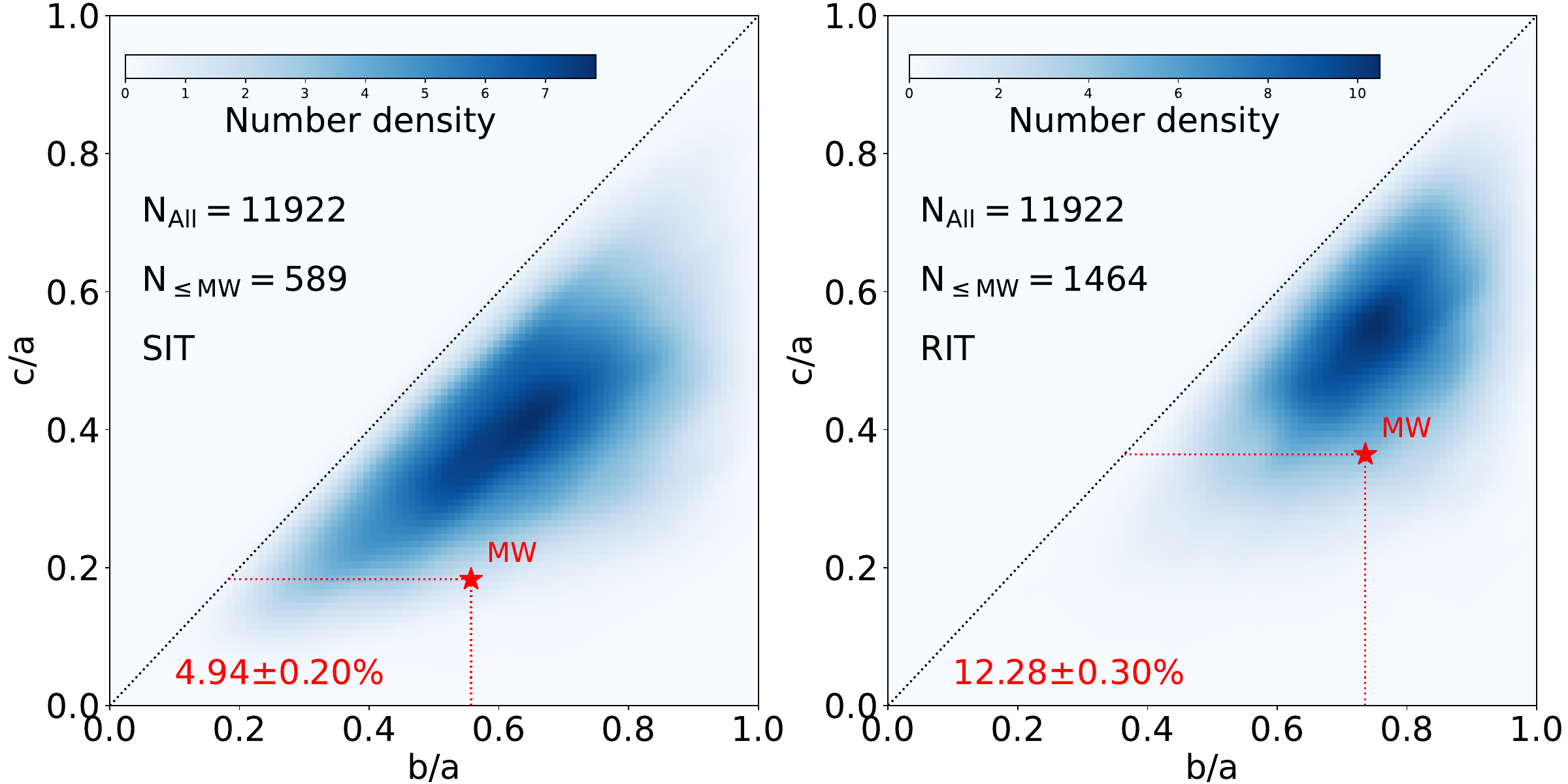}
\caption{Joint distribution of $b/a$ and $c/a$ for the 11 most massive satellites of the 11,922 simulated MW analogues. Colour indicates the smoothed number density. The left and right panels show SIT and RIT, respectively. The red star marks the MW, while the red dotted lines bound systems for which both ratios are no larger than the MW values. The quoted fraction is $N_{\leq\mathrm{MW}}/N_{\rm all}$, with binomial uncertainties.}
\label{fig:f3}
\end{figure}

\begin{figure}[htbp]
\includegraphics[width=0.48\textwidth]{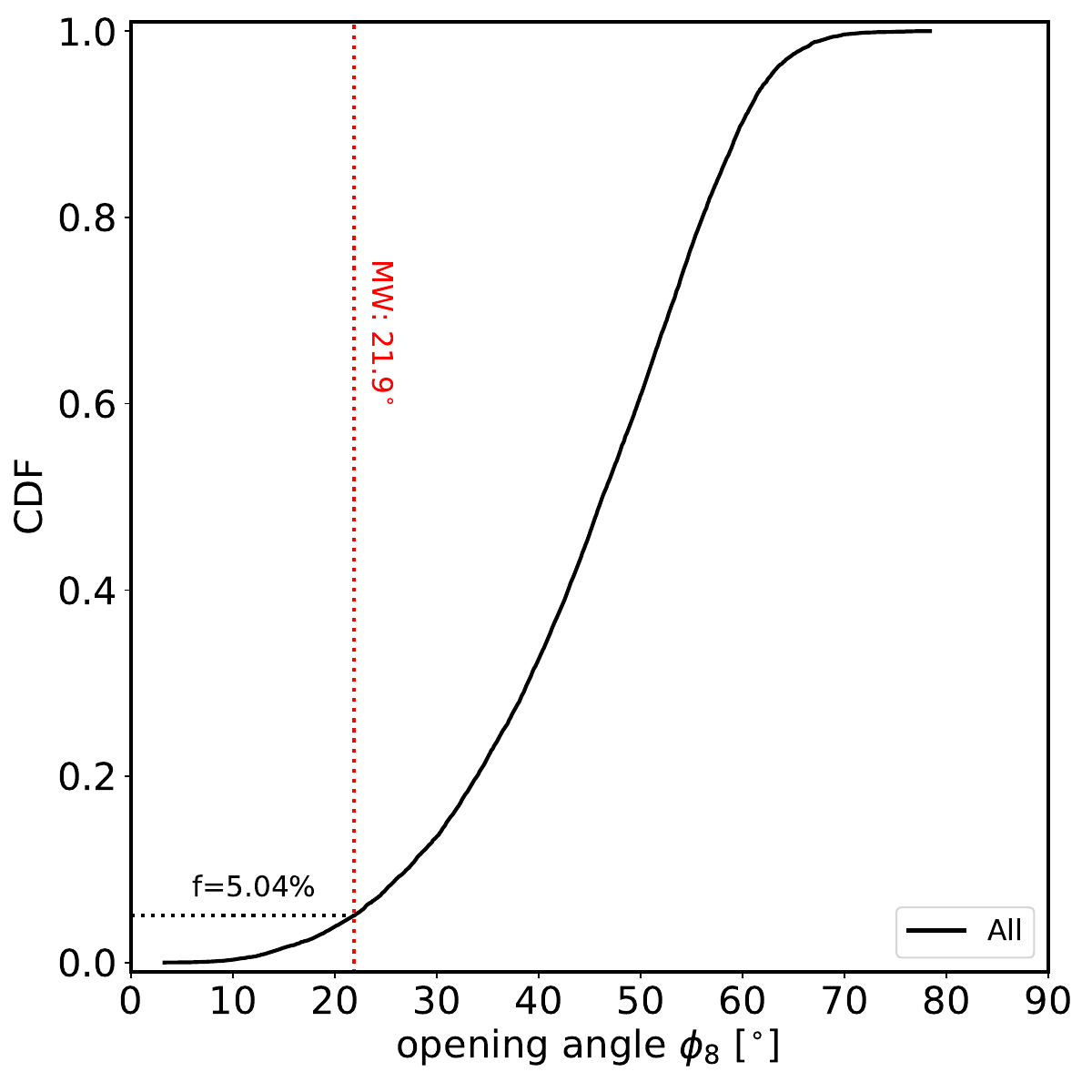}
\includegraphics[width=0.48\textwidth]{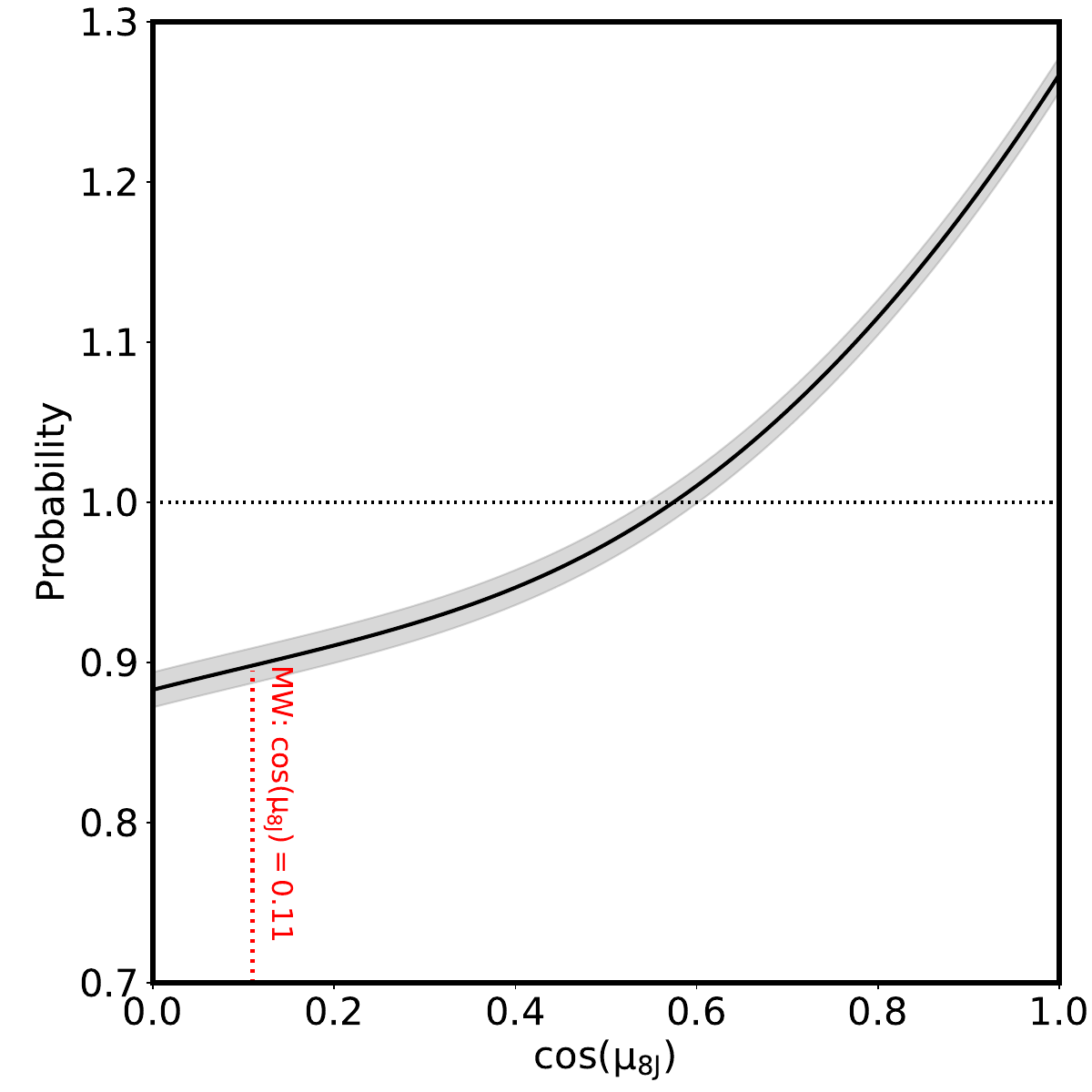}
\caption{\textbf{Left:} Cumulative distribution of the minimum eight-orbit opening angle $\phi_8$ for the simulated hosts. The red dashed line marks the MW value of $21.9^\circ$, below which $5.04\%$ of the simulated systems lie. \textbf{Right:} Probability density of $\mu_{8J}=|\hat{\boldsymbol{J}}^8_{\rm mean}\cdot\hat{\boldsymbol{J}}_{\rm host}|$, normalized by the isotropic expectation. The grey band indicates the estimated uncertainty, the dotted line denotes isotropy, and the red dashed line marks the MW value of 0.11.}
\label{fig:opening_angle}
\end{figure}

The joint shape distributions are shown in Fig.~\ref{fig:f3}. The SIT distribution is centred on moderately flattened systems, whereas RIT shifts the population towards larger axis ratios. The MW lies in the low-$c/a$ tail in both panels, but is considerably more extreme for SIT. In total, 589 SIT systems ($4.94\pm0.20\%$) and 1,464 RIT systems ($12.28\pm0.30\%$) have both axis ratios no larger than the corresponding MW values. The difference of a factor of about 2.5 shows that the inferred rarity depends strongly on whether the radial information is retained.

The cumulative distribution of $\phi_8$ is shown in the left panel of Fig.~\ref{fig:opening_angle}. Only $5.04\%$ of the simulated systems satisfy $\phi_8\leq21.9^\circ$. Because this statistic is based on orbital axes and does not depend on the instantaneous spatial shape, its small tail fraction is not a second measurement of $c/a$.

The right panel compares the $\mu_{8J}$ distribution with the isotropic expectation. The simulated orbital-plane axes show a mild preference for alignment with the host spin, most clearly at large $\mu_{8J}$. By contrast, the MW value of $\mu_{8J}=0.11$ describes an almost polar orbital plane and falls in a region that is mildly underrepresented relative to isotropy. The MW is therefore unusual in two distinct respects: its eight most coplanar orbital axes are tightly clustered, and their common axis is nearly perpendicular to the host spin.

\subsection{Environmental Dependence of Satellite Planarity and Orbital-Pole Coherence}

We consider the environment on two scales. On Local Group scales, a host belongs to the ``Pair'' sample if another MW-mass host lies within a three-dimensional distance of $1.5\,\mathrm{Mpc}$. This intentionally broad limit includes the present MW--M31 separation \citep{2012AJ....144....4M}, but is not designed to reproduce all of the Local Group's dynamical or isolation properties. 
Galaxy pairs resembling the Local Group have previously been shown to exhibit
satellite and large-scale alignment signals that depend on their surrounding
filamentary environment
\citep{2019MNRAS.484.4325W,2025ApJ...994..207W}.
The resulting Pair sample contains 3,992 hosts. On larger scales, we use both the continuous velocity-shear eigenvalues and the four web classes defined in Section~\ref{sec:data_method}.

\begin{table}[htbp]
\centering
\caption{%
Velocity-shear eigenvalues and eigenvectors at the Local Group position from ref.~\cite{2015MNRAS.452.1052L}. The eigenvectors are expressed in the Galactocentric Cartesian frame. The lower rows give acute alignment angles and absolute cosines for the MW spin, $\hat{\boldsymbol J}_{\rm MW}=(0,0,-1)$, and the folded mean axis of the eight most coplanar satellite orbits, $\hat{\boldsymbol J}^{8}_{\rm mean}=(-0.994,-0.021,-0.106)$.
}
\label{tab:lss}
\setlength{\tabcolsep}{10pt}
\renewcommand{\arraystretch}{1.3}
\begin{tabular}{lccc}
\hline\hline
Eigenvalue $\lambda_i$ & $\lambda_1$ & $\lambda_2$ & $\lambda_3$ \\
\hline
 & $0.148\pm0.038$ & $0.051\pm0.039$ & $-0.160\pm0.033$ \\
\hline
eigenvectors $\boldsymbol{e}_i$ & $\boldsymbol{e}_1$ & $\boldsymbol{e}_2$ & $\boldsymbol{e}_3$ \\
\hline
$x$-component & $-0.866$ & $-0.311$ & $0.392$ \\
$y$-component & $-0.451$ & $0.826$ & $-0.338$ \\
$z$-component & $0.219$ & $0.469$ & $0.855$ \\
\hline\hline
$\angle(\boldsymbol{J}_{\rm MW},\,\boldsymbol{e}_i)$ & $77.4^\circ$ & $62.0^\circ$ & $31.2^\circ$ \\
$|\cos\mu|$ & $0.219$ & $0.469$ & $0.855$ \\
\hline
$\angle(\hat{\boldsymbol{J}}^8_{\rm mean},\,\boldsymbol{e}_i)$ & $32.2^\circ$ & $75.9^\circ$ & $61.7^\circ$ \\
$|\cos\mu|$ & $0.847$ & $0.243$ & $0.474$ \\
\hline\hline
\end{tabular}
\end{table}

Table~\ref{tab:lss} summarizes the Local Group reconstruction. The two positive eigenvalues classify the region as a filament under the zero-threshold convention, with $\boldsymbol e_3$ directed along the filament spine. The MW spin lies relatively close to $\boldsymbol e_3$ ($\mu_{J3}=0.855$) and is nearly perpendicular to $\boldsymbol e_1$ ($\mu_{J1}=0.219$). In contrast, the satellite orbital-plane axis is close to $\boldsymbol e_1$ ($\mu_{81}=0.847$). Together with $\mu_{8J}=0.11$, these measurements show that the disc and the satellite orbital plane select different directions in the same local shear frame. This geometry motivates a statistical comparison, but does not by itself demonstrate that the present-day shear established either orientation. It may instead record different aspects, or different epochs, of the MW's assembly history: the stellar disc and satellite system could have acquired their present angular-momentum directions through distinct episodes of anisotropic accretion and subsequent dynamical evolution \citep{2017MNRAS.468L.123W,2018MNRAS.473.1562W,2018ApJ...859..115W,2022MNRAS.516.4576D,2026A&A...706A.193Z}.

\begin{figure}[htbp]
\centering
\includegraphics[width=0.98\textwidth]{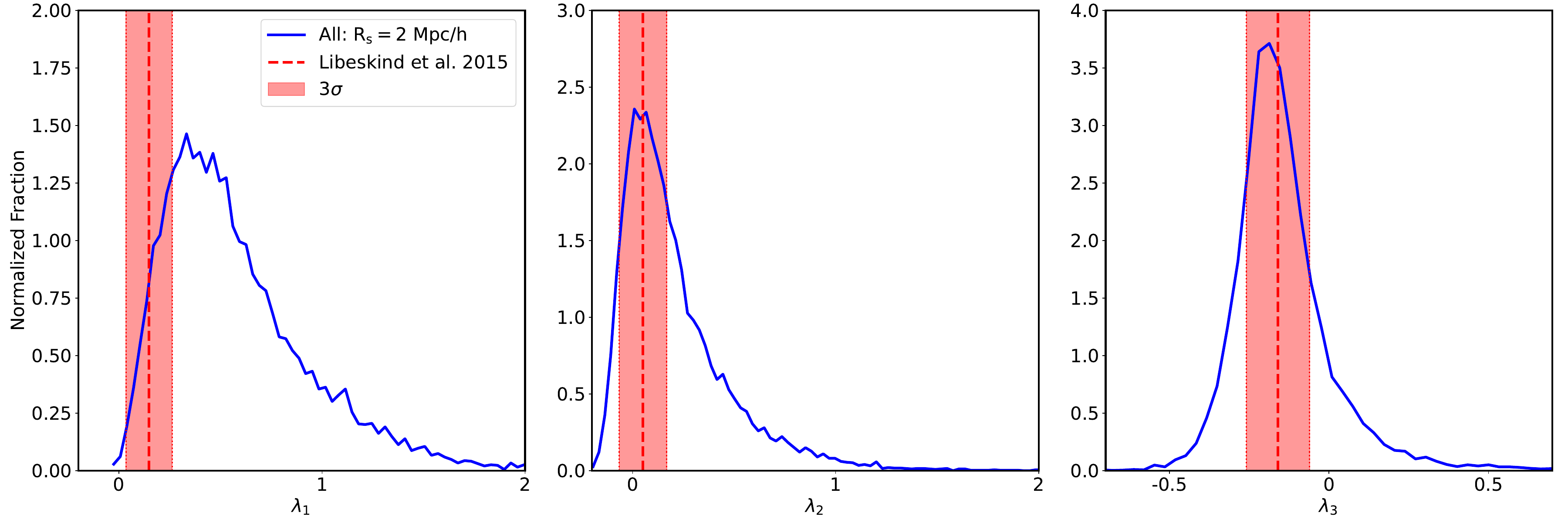}
\caption{
Probability densities of the velocity-shear eigenvalues $\lambda_1$, $\lambda_2$, and $\lambda_3$ for the simulated MW-mass hosts, using $R_s=2\,h^{-1}\mathrm{Mpc}$. The red dashed lines and shaded bands mark the Local Group values and their $3\sigma$ intervals from ref.~\cite{2015MNRAS.452.1052L}.
}
\label{fig:f5}
\end{figure}

Following common practice in cosmic-web analyses, we adopt $R_s=2 \ h^{-1}\mathrm{Mpc}$ as the fiducial smoothing scale. Figure~\ref{fig:f5} locates the reconstructed eigenvalues within the simulated distributions at this scale. All three Local Group intervals overlap well-sampled regions of the TNG300-1 distributions, although $\lambda_1$ lies below the mode of the simulated sample. Applying the three $3\sigma$ cuts simultaneously selects 2,049 LSS-matched hosts.

\begin{figure}[htbp]
\centering
\includegraphics[width=0.98\textwidth]{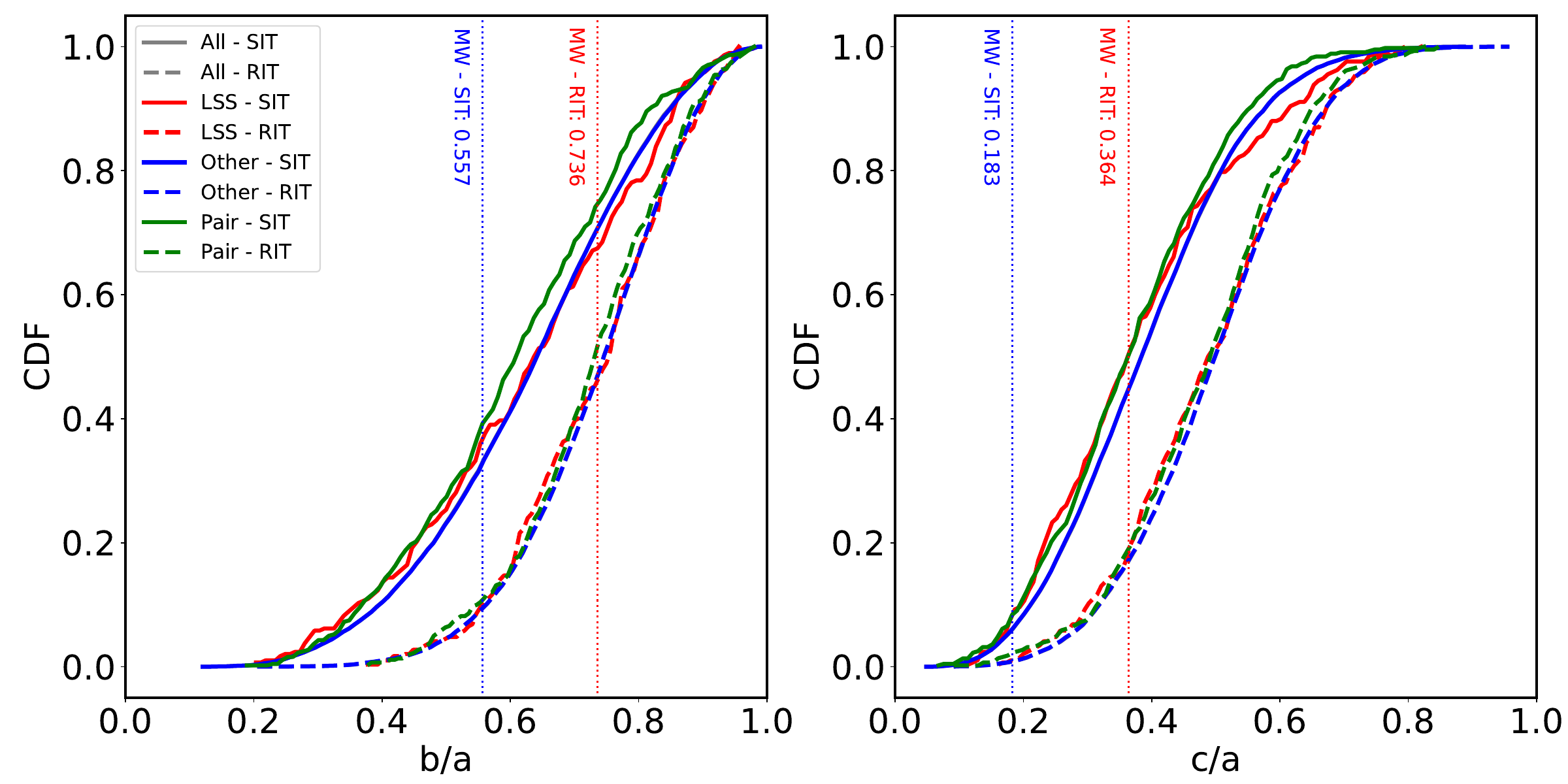}
\caption{Cumulative distributions of $b/a$ (left) and $c/a$ (right). Solid and dashed curves show SIT and RIT, respectively. Grey, red, blue, and green denote the full, LSS-matched, non-LSS-matched, and Pair samples. The vertical blue and red dotted lines mark the MW values for SIT and RIT.}
\label{fig:f6}
\end{figure}

Figure~\ref{fig:f6} compares the one-dimensional shape distributions. The full sample and the three environmental selections produce largely overlapping curves for both tensor definitions. There is a small shift in SIT $b/a$ for the Pair sample and minor deviations for the LSS-matched sample, but neither selection shifts both $b/a$ and $c/a$ systematically towards the MW values. We emphasize that RIT is a reduced spatial tensor, not a rotation statistic; the similarity of its curves refers to angular flattening, not orbital coherence.

\begin{figure}[htbp]
\centering
\includegraphics[width=0.98\textwidth]{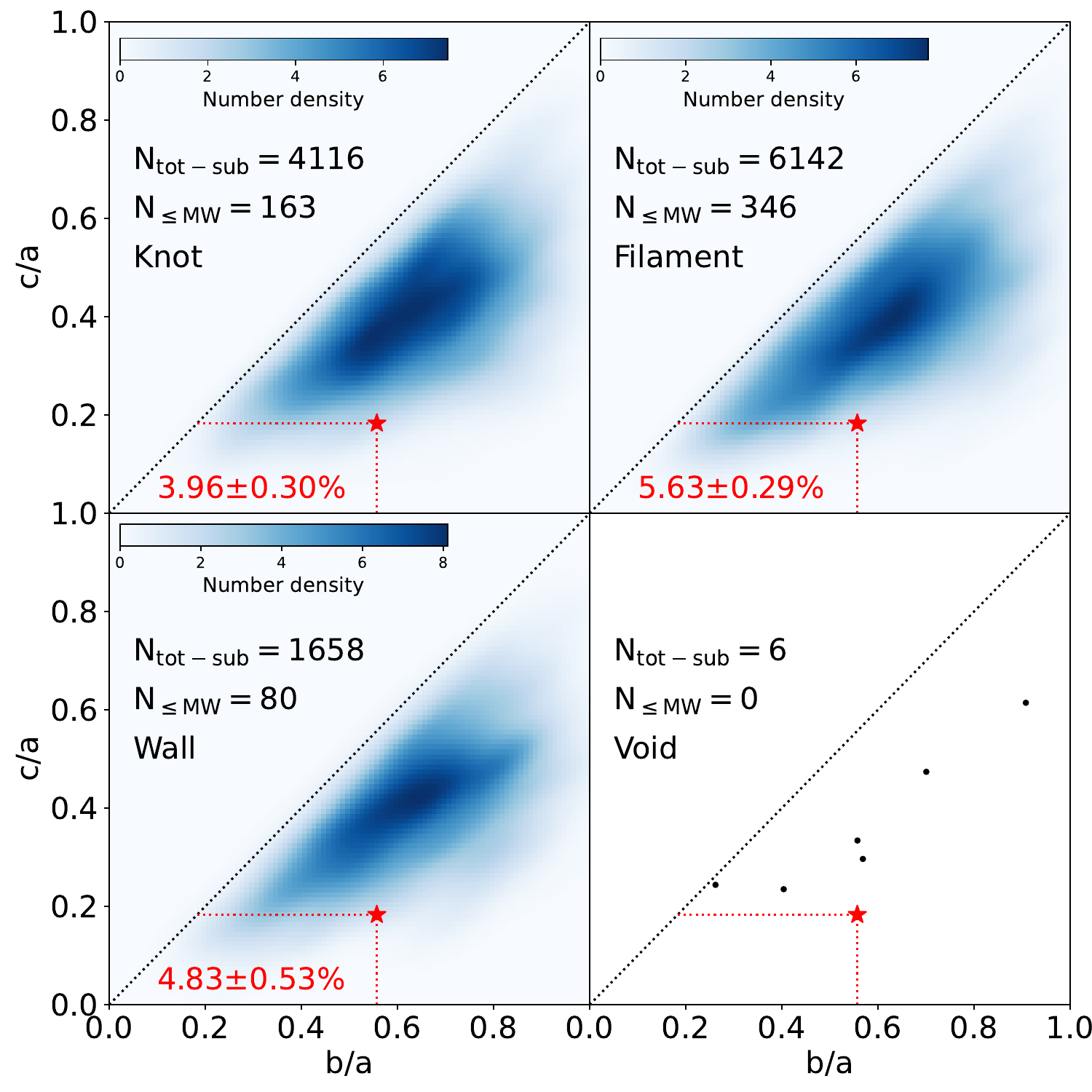}
\caption{
SIT shape distributions in knots, filaments, sheets (labelled ``Wall''), and voids. Colour indicates number density, the red star marks the MW, and the red dotted lines delimit systems for which both $b/a$ and $c/a$ are no larger than the MW values. The red percentage gives the corresponding joint tail fraction and its binomial uncertainty.
}
\label{fig:f7}
\end{figure}

\begin{figure}[htbp]
\centering
\includegraphics[width=0.98\textwidth]{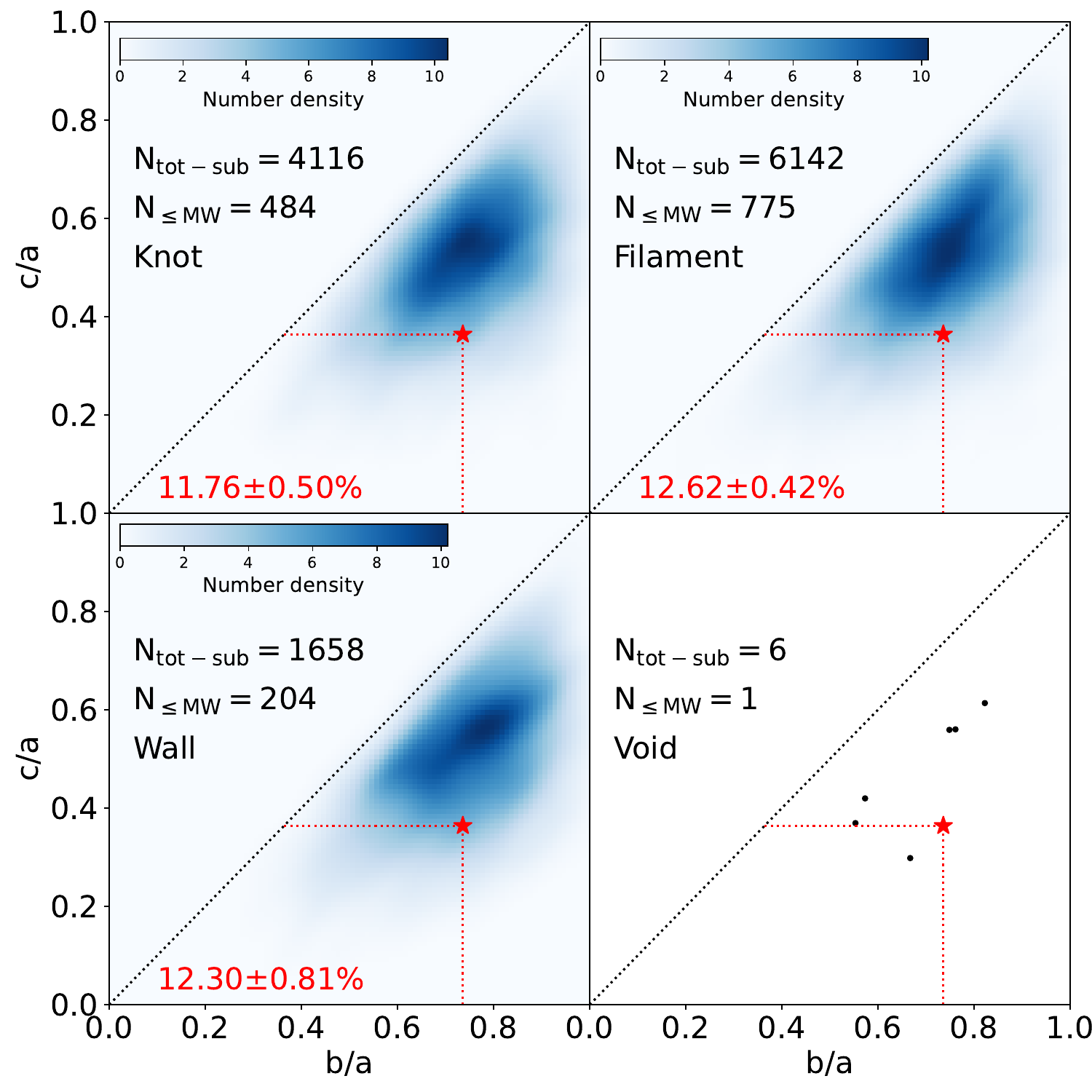}
\caption{As Fig.~\ref{fig:f7}, but for the reduced inertia tensor (RIT).}
\label{fig:f8}
\end{figure}

Figures~\ref{fig:f7} and \ref{fig:f8} separate the same comparison by web class. The knot, filament, and sheet samples occupy similar loci in the $b/a$--$c/a$ plane. Their SIT joint tail fractions are $3.96\%$, $5.63\%$, and $4.83\%$, while their RIT fractions are $11.76\%$, $12.62\%$, and $12.30\%$, respectively. These differences are small compared with the difference between SIT and RIT, and they do not vary monotonically with the number of collapsing directions. Because only six selected hosts reside in voids, the void fractions ($0/6$ for SIT and $1/6$ for RIT) are not statistically informative. Observational reconstructions of the local velocity and density fields place the Local Group in a filament extending towards the Virgo Cluster, embedded in the Local Sheet and adjacent to the Local Void \citep{2013AJ....146...69C,2013AJ....146...86T,2015MNRAS.452.1052L}. The similar tail fractions found here nevertheless indicate that this particular cosmic-web setting does not, by itself, substantially enhance the incidence of MW-like satellite flattening.

\begin{figure}[htbp]
\centering
\includegraphics[width=0.4\textwidth]{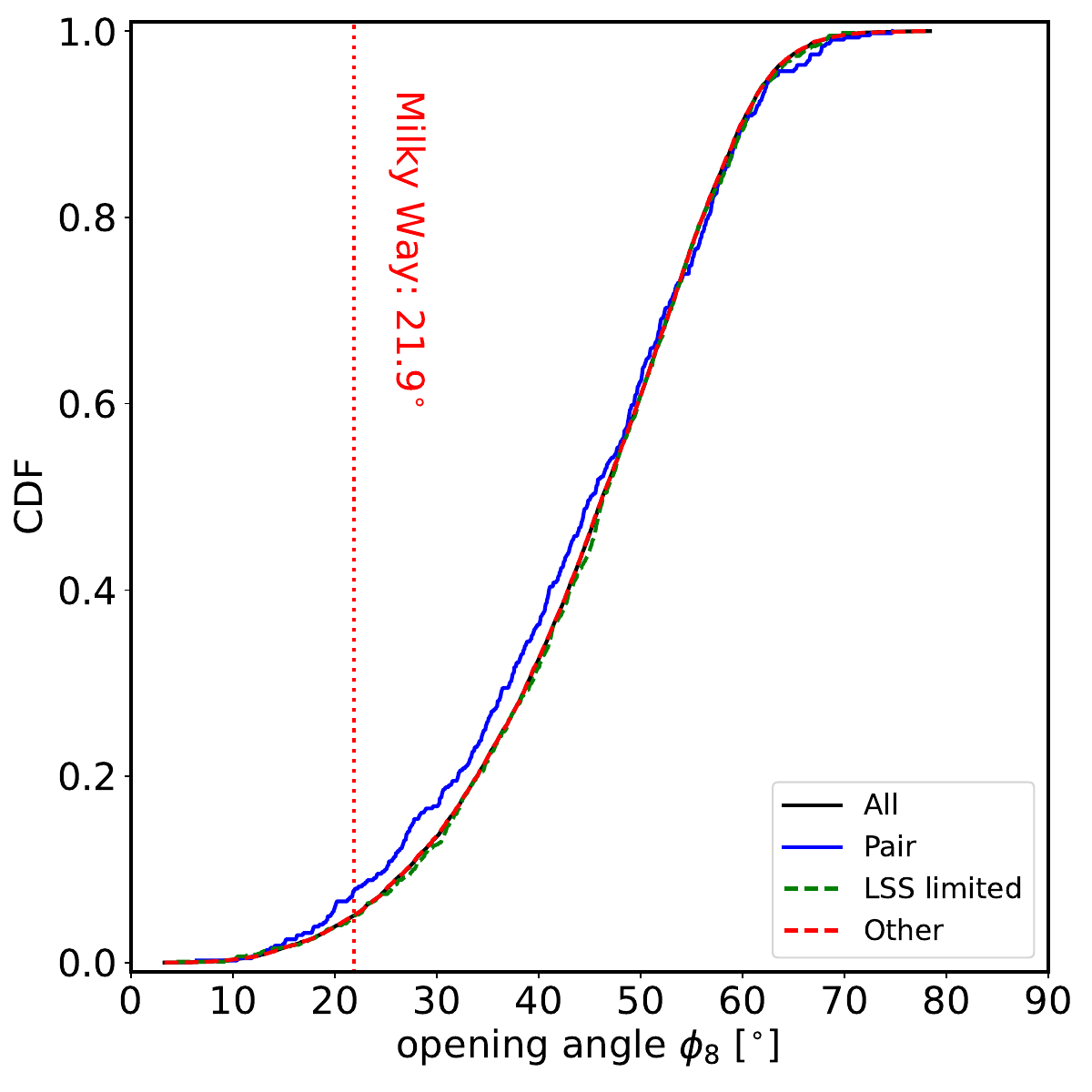}
\qquad
\includegraphics[width=0.4\textwidth]{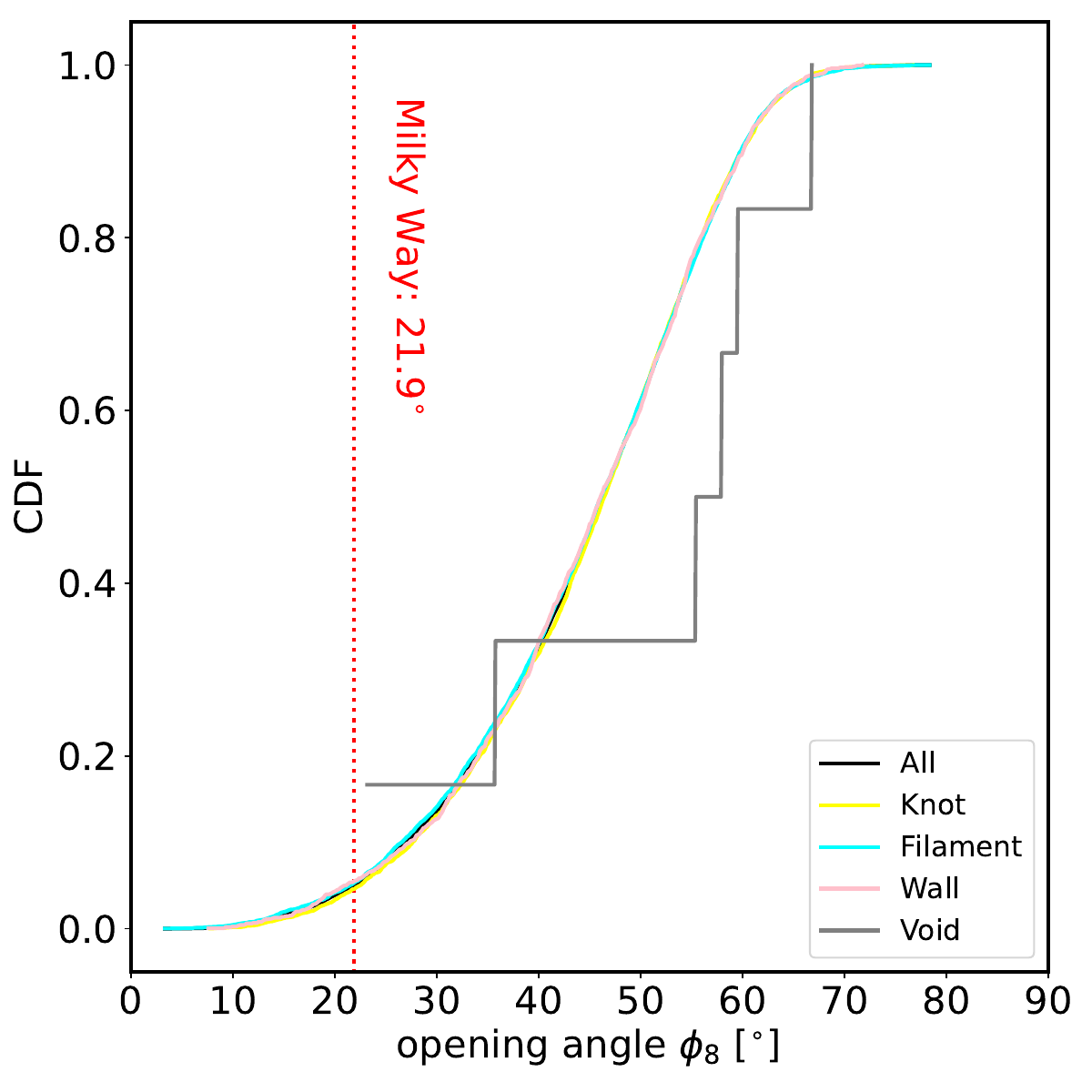}
\caption{
Cumulative distributions of the minimum opening angle $\phi_8$ for the eight most coplanar satellite orbits. \textbf{Left:} full, Pair, LSS-matched, and non-LSS-matched samples. \textbf{Right:} full sample and the four web classes. The vertical red dotted line marks the MW value of $21.9^\circ$.
}
\label{fig:opening_angle_cdf}
\end{figure}
\FloatBarrier

\begin{table}[htbp]
\caption{Sample sizes and fractions of systems with both $b/a$ and $c/a$ no larger than the MW values. The knot, filament, sheet, and void counts sum to the full sample.}
\label{tab:environmental_fractions_p}
\begin{center}
\renewcommand{\arraystretch}{1.2}
\setlength{\tabcolsep}{10pt}
\begin{tabular}{lccc}
\hline\hline
Sample & $N$ & $f_{\mathrm{SIT}}$ (\%) & $f_{\mathrm{RIT}}$ (\%) \\ 
\hline
All         & 11922 & 4.94 & 12.28 \\ 
Pair        & 3992  & 4.96 & 12.36 \\ 
LSS matched & 2049  & 4.93 & 12.15 \\
Knot        & 4116  & 3.96 & 11.76 \\ 
Filament    & 6142  & 5.63 & 12.62 \\ 
Sheet       & 1658  & 4.83 & 12.30 \\ 
Void        & 6     & 0.00 & 16.67 \\  
\hline\hline
\end{tabular}
\end{center}
\end{table}

Table~\ref{tab:environmental_fractions_p} quantifies these trends. Relative to the full sample, the Pair and LSS-matched selections alter the tail fractions by less than $0.2$ percentage points. Among the three well-sampled web classes, the largest differences are only $1.67$ percentage points for SIT and $0.86$ percentage points for RIT. The incidence of flattened satellite systems therefore depends only weakly on these environmental selections.

The opening-angle distributions in Fig.~\ref{fig:opening_angle_cdf} give the same qualitative result for orbital planes. The full, Pair, LSS-matched, and complementary curves remain close to one another, as do those for knots, filaments, and sheets. At the MW value, the cumulative fractions are only a few per cent. The six-object void distribution is visibly discrete and cannot be used to infer a void-specific trend. Neither present-day pair membership nor web class therefore explains the small MW value of $\phi_8$.

We next combine the individual MW-like properties in a cumulative selection. We first apply the two axis-ratio cuts, and then require an opening angle at least as small as the observed MW value, $\phi_8\leq21.9^\circ$. We subsequently require the orbital-plane axis to be at least as polar relative to the host spin as in the MW, $\mu_{8J}\leq0.11$, and finally apply the LSS-matched selection. Table~\ref{tab:combined_criteria} lists both the number and fraction of systems remaining after each successive cut.

\begin{table}[htbp]
\caption{Numbers and fractions of simulated systems satisfying successive MW-like criteria. The opening-angle criterion is the observed MW value, $\phi_8\leq21.9^\circ$; the next step requires an orbital-plane--host-spin orientation at least as polar as that of the MW, $\mu_{8J}\leq0.11$; and the final step requires all three LSS eigenvalues to fall within the Local Group $3\sigma$ intervals.}
\begin{center}
\begin{tabular}{lcccc}
\hline\hline
Criterion & $N_{\rm SIT}$ & $f_{\rm SIT}$ (\%) & $N_{\rm RIT}$ & $f_{\rm RIT}$ (\%) \\ \hline
$(b/a,c/a)$ & 589 & 4.94 & 1464 & 12.28 \\
$+(\phi_8\leq21.9^\circ)$ & 219 & 1.84 & 368 & 3.09 \\
$+(\mu_{8J}\leq0.11)$ & 35 & 0.29 & 113 & 0.95 \\
$+$ LSS matched & 0 & 0 & 7 & 0.06 \\ 
\hline\hline
\end{tabular}
\end{center}
\label{tab:combined_criteria}
\end{table}

The shape cuts retain 589 SIT systems and 1,464 RIT systems. Adding the observed opening-angle requirement leaves 219 and 368 systems, respectively. The orbital-plane--host-spin orientation cut further reduces these samples to 35 SIT systems and 113 RIT systems. After the LSS cut, 0 SIT systems and only 7 RIT systems satisfy all four criteria, corresponding to the tabulated fractions of $0.0\%$ and $0.06\%$. These small fractions quantify the adopted multivariate selection, but should not be interpreted as a rejection significance for $\Lambda$CDM because the criteria are correlated and were chosen with prior knowledge of the MW.

\FloatBarrier

\begin{figure}[htbp]
\centering
\includegraphics[width=0.45\textwidth]{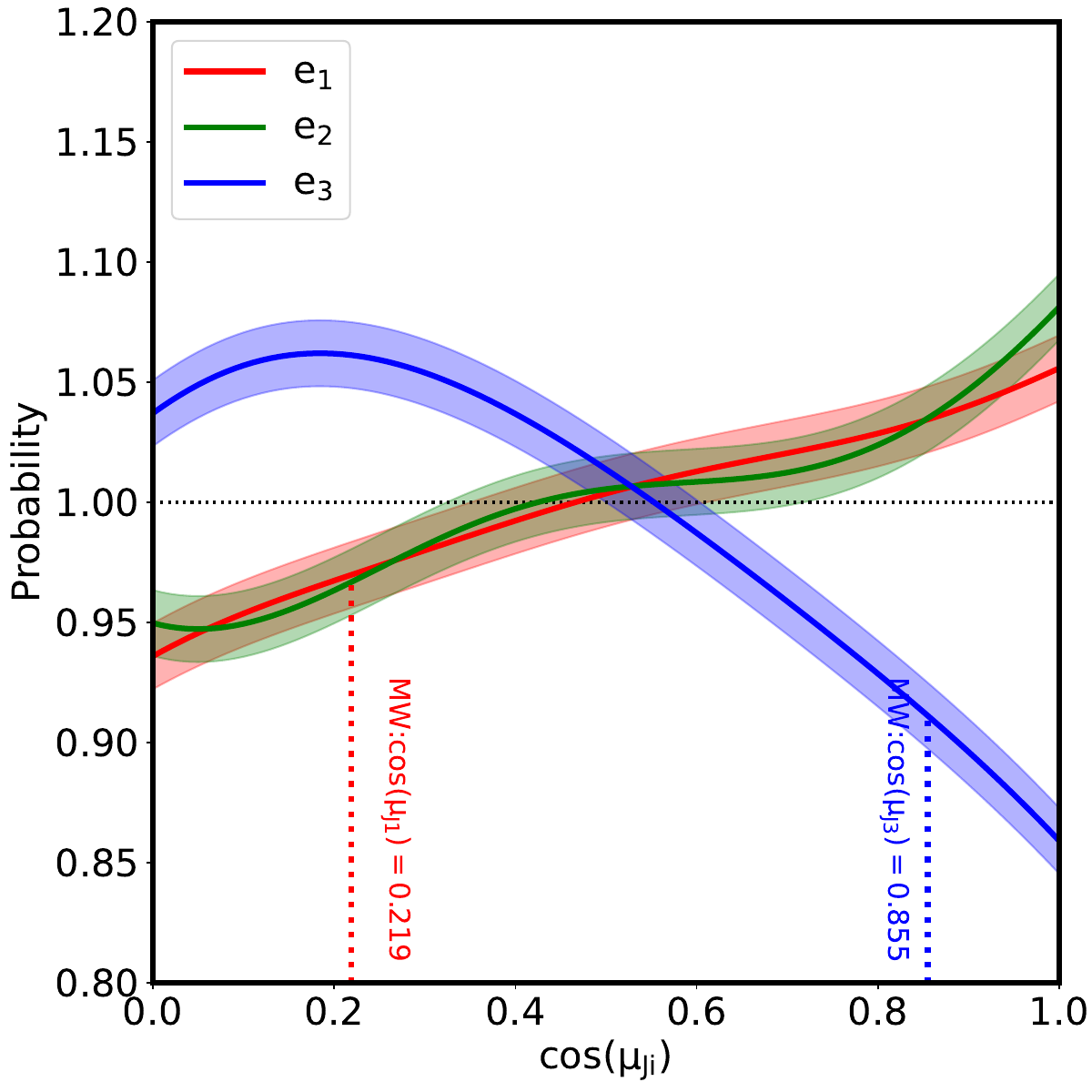}
\includegraphics[width=0.45\textwidth]{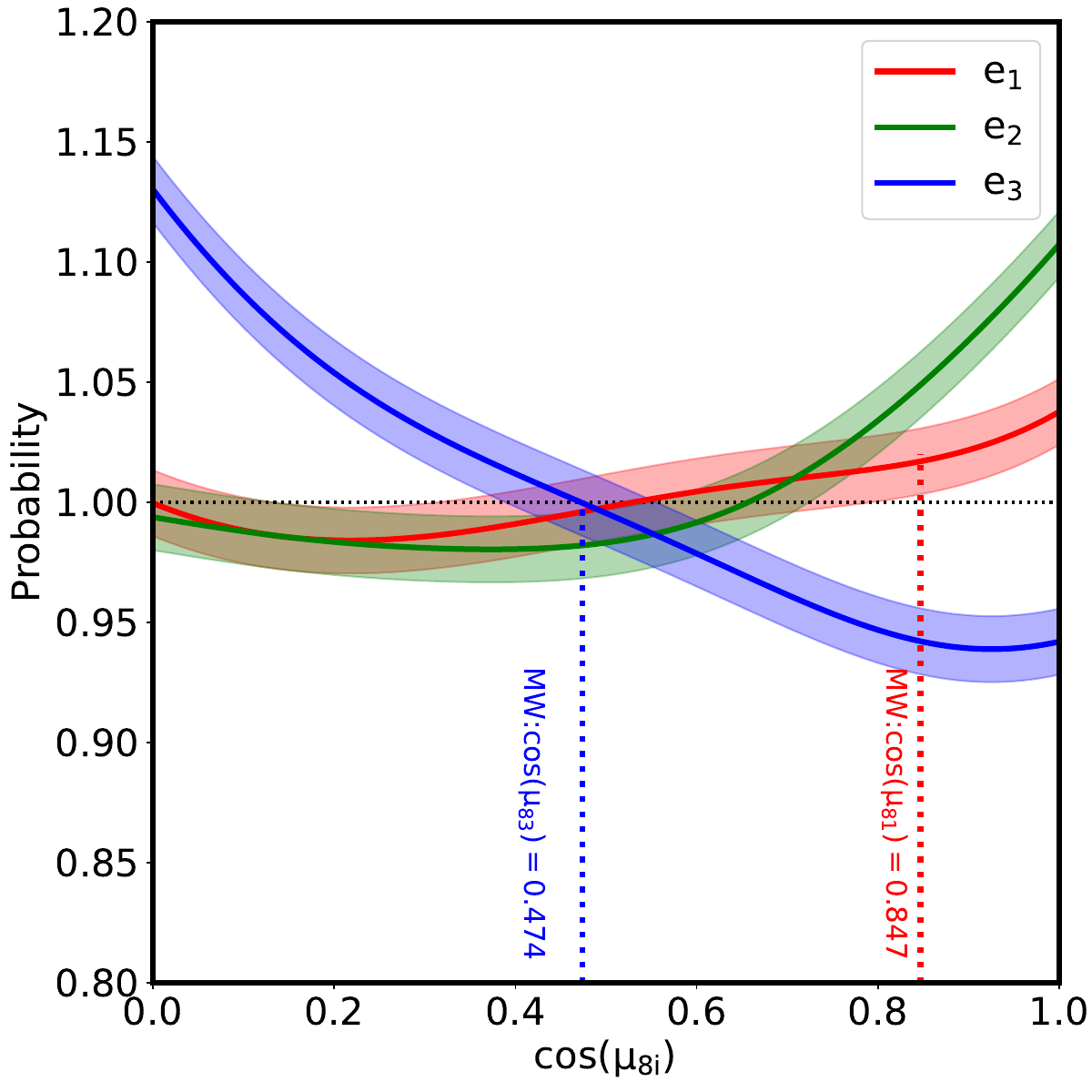}
\caption{Probability densities, normalized by the isotropic expectation, for alignments with the velocity-shear eigenvectors. \textbf{Left:} host spin, $|\hat{\boldsymbol J}_{\rm host}\cdot\boldsymbol e_i|$. \textbf{Right:} folded satellite orbital-plane axis, $|\hat{\boldsymbol J}^{8}_{\rm mean}\cdot\boldsymbol e_i|$. The red, green, and blue curves correspond to $\boldsymbol e_1$, $\boldsymbol e_2$, and $\boldsymbol e_3$, respectively; shaded bands show the estimated uncertainties. Coloured vertical dashed lines mark the corresponding MW values from Table~\ref{tab:lss}.}
\label{fig:alignment_simu}
\end{figure}
\FloatBarrier

The orientation distributions retain information that is lost when the environment is reduced to a web-class label. In the left panel of Fig.~\ref{fig:alignment_simu}, simulated host spins show a weak avoidance of $\boldsymbol e_3$ and a corresponding excess of alignment with $\boldsymbol e_1$ and $\boldsymbol e_2$ at large alignment cosines. With $\mu_{J3}=0.855$, the MW spin lies on the underrepresented side of the $\boldsymbol e_3$ relation. In the right panel, the satellite orbital-plane axes also avoid $\boldsymbol e_3$ and show a preference for alignment with $\boldsymbol e_1$ or $\boldsymbol e_2$. The MW value of $\mu_{81}=0.847$ is qualitatively consistent with this tendency. The cosmic web can therefore influence the orientations of satellite systems even though it has little effect on the one-point distributions of their flattening and opening angles. This distinction is consistent with anisotropic accretion organizing satellite orbits without producing a strong environmental dependence in the present-day incidence of flattened or orbitally coherent satellite configurations \citep{2015MNRAS.452.1052L,2024ApJ...965..154G}.


\section{Conclusions and Discussion}
\label{sec:con_dis}

We have compared the 11 classical MW satellites with the 11 most massive satellites around 11,922 MW-mass hosts in TNG300-1, applying the same spatial and orbital-plane statistics to both. We considered two measures of environment: the presence of another MW-mass host within $1.5\,\mathrm{Mpc}$ and the velocity-shear field smoothed on $2\,h^{-1}\mathrm{Mpc}$ scales. Our main results are as follows.

\begin{itemize}
\item The MW satellite system is unusually flattened, although the inferred tail fraction depends on the spatial statistic. Requiring both $b/a$ and $c/a$ to be no larger than the MW values selects $4.94\%$ of the simulated systems for SIT and $12.28\%$ for RIT. The larger RIT fraction shows that part of the extreme SIT flattening is due to the radial leverage of the outer satellites.

\item The axes of the eight most coplanar MW satellite orbits lie within $\phi_8=21.9^\circ$, a value reached by $5.04\%$ of the simulated systems. Because the statistic folds antipodal poles, this result indicates orbital-plane coherence rather than eight-member co-rotation; Sculptor is the counter-rotating member of the MW subset.

\item Neither selection for Local Group-like pairs nor matching the three Local Group shear eigenvalues appreciably increases the incidence of MW-like flattening. The knot, filament, and sheet samples also have similar shape and opening-angle distributions. With only six objects, the void sample does not support an independent conclusion.

\item The lack of a strong environmental trend in incidence does not imply that the cosmic web is dynamically unimportant. Simulated host spins and satellite orbital-plane axes are oriented anisotropically relative to the velocity-shear frame. In particular, the satellite axis tends to avoid the slowest-collapse direction $\boldsymbol e_3$ and to align with $\boldsymbol e_1$ or $\boldsymbol e_2$. The MW satellite axis follows the $\boldsymbol e_1$ tendency, whereas the MW disc spin lies unusually close to $\boldsymbol e_3$.
\end{itemize}

Correlated and anisotropic infall is a generic feature of $\Lambda$CDM and can organize
satellite positions and orbits
\citep{2011MNRAS.411.1525L,2015MNRAS.452.1052L,2020ApJ...900..129W,
2022MNRAS.516.4576D,2024ApJ...965..154G}.
The present-day web class and the three shear eigenvalues, however, do not uniquely
specify the accretion history of an individual halo. Group infall, recent massive accretion
events, halo torques, and the orbital phases of a small number of distant satellites can all
modify the configuration observed at a given epoch
\citep{2015MNRAS.449.2576C,2019MNRAS.488.1166S,2023NatAs...7..481S}.
Previous studies have also shown that Local Group-like galaxy pairs and their satellite
systems exhibit statistically significant alignments with the surrounding filamentary
environment
\citep{2019MNRAS.484.4325W,2025ApJ...994..207W}.
Such correlations indicate that the large-scale environment can imprint preferred directions
on galaxy pairs and satellite populations, but they do not imply that a particular filamentary
configuration will generically produce a thin and orbitally coherent satellite plane.
The weak environmental dependence found here is therefore consistent with a cosmic-web
imprint on preferred directions, while disfavouring a simple deterministic picture in which
either a filament or a Local Group-like pair necessarily gives rise to an MW-like satellite plane.

The present configuration may also be transient. Simulations that follow MW-mass systems
through multiple snapshots find purely spatial MW-like alignments to be short-lived:
less than $1\,\mathrm{Gyr}$ in EAGLE, typically below $500\,\mathrm{Myr}$ in FIRE,
and no more than a few hundred Myr in a recent TNG50 analysis
\citep{2019MNRAS.488.1166S,2021MNRAS.504.1379S,2026Sawala}.
The importance of temporal evolution is also illustrated by \citet{2023ApJ...954..128X},
who identified an MW-like satellite plane in TNG50 and found that most of its members
could be related to the anisotropic large-scale environment, whereas several satellites
occupied the appropriate positions and velocities only at $z=0$.
By contrast, the orbital planes, or the earlier spatial coherence of a fixed subset of satellites,
can persist for several Gyr. A single $z=0$ snapshot therefore cannot distinguish between
a transient geometric alignment and a longer-lived, dynamically related satellite population.
Following the selected satellites along their merger trees would establish whether the
TNG300-1 analogues are brief alignments or longer-lived correlated populations, and would
separate changes caused by membership turnover from the dispersal of a fixed set of satellites.

More broadly, the present orientations need not have a single origin. The stellar-disc spin
can evolve through gas accretion and merger-driven torques, whereas the satellites may inherit
correlated positions and angular momenta from their larger-scale Lagrangian environment
\citep{2011MNRAS.411.1525L,2017MNRAS.468L.123W,2018MNRAS.473.1562W,
2024ApJ...965..154G,2026A&A...706A.193Z}.
After infall, nonlinear evolution within the host halo---including dynamical friction, tidal
disruption, orbital precession, and interactions with the disc and other satellites---can further
reshape the satellite distribution, while baryonic physics can affect which satellites survive
to the present day \citep{2017MNRAS.466.3119A}.
The unusual MW configuration may therefore reflect the combined effects of disc formation,
the large-scale origin of the surviving satellites, and their subsequent internal evolution.
Future high-resolution simulations that follow these processes self-consistently can quantify
their relative contributions.

Several limitations should be kept in mind. TNG300-1 provides the statistical volume required
for this analysis, but only marginally resolves the least massive luminous satellites, and ranking
by stellar mass is not identical to an observational selection by luminosity. The MW benchmark
uses central phase-space values without marginalizing over proper-motion and distance
uncertainties. The zero eigenvalue threshold and the $2\,h^{-1}\mathrm{Mpc}$ smoothing scale
are specific choices, while the present-day pair criterion does not constrain the time-dependent
orbit of the companion. Finally, the reported tail fractions are a posteriori descriptors rather
than formal $p$-values. Repeating the matched analysis in higher-resolution and constrained
simulations, while propagating the MW measurement covariance, would provide an important
test of the robustness of these conclusions.

\acknowledgments
P.W thank Sun YY and Wang QX for valuable discussions and part of the double-blind calculations.
Y.L., J.C., and X.K. acknowledge financial support from the NSFC grant (No. 12533007).
Y.L.  acknowledges financial support from the NSFC grant (No. 12442061).
P.W. acknowledges financial support from the NSFC grant (Nos. 12595312, 12473009) and the Shanghai Rising-Star Program (No. 24QA2711100). This work is supported by the China Manned Space Program through grant No. CMS-CSST-2025-A03.
J.C acknowledges financial support from the NSFC grants (Nos.12573024, and 12233005), the National Key Research and Development Program of China (2023YFA1608100), and the China Manned Space Program with grant No. CMS-CSST- 2025-A20
S.S acknowledges financial support from the NSFC grants (Nos. 11988101, 12273053).

\bibliographystyle{JHEP}
\bibliography{main}
\end{document}